 \documentclass[sigconf]{acmart}

\AtBeginDocument{%
  \providecommand\BibTeX{{%
    \normalfont B\kern-0.5em{\scshape i\kern-0.25em b}\kern-0.8em\TeX}}}

\copyrightyear{2022}
\acmYear{2022}
\setcopyright{acmlicensed}\acmConference[CHI '22]{CHI Conference on Human Factors in Computing Systems}{April 29-May 5, 2022}{New Orleans, LA, USA}
\acmBooktitle{CHI Conference on Human Factors in Computing Systems (CHI '22), April 29-May 5, 2022, New Orleans, LA, USA}
\acmPrice{15.00}
\acmDOI{10.1145/3491102.3517728}
\acmISBN{978-1-4503-9157-3/22/04}

\usepackage{soul}

\begin{document}

\title{The Dark Side of Perceptual Manipulations in Virtual Reality}

\author{Wen-Jie Tseng}
\affiliation{%
  \institution{LTCI, Telecom Paris, IP Paris}
  \city{Palaiseau}
  \state{}
  \country{France}
}
\email{wen-jie.tseng@telecom-paris.fr}

\author{Elise Bonnail}
\affiliation{%
  \institution{LTCI, Telecom Paris, IP Paris}
  \city{Palaiseau}
  \state{}
  \country{France}
}
\email{elise.bonnail@telecom-paris.fr}

\author{Mark McGill}
\affiliation{%
  \institution{University of Glasgow}
  \city{Glasgow}
  \state{Scotland}
  \country{UK}
}
\email{mark.mcgill@glasgow.ac.uk}

\author{Mohamed Khamis}
\affiliation{%
  \institution{University of Glasgow}
  \city{Glasgow}
  \state{Scotland}
  \country{UK}
}
\email{mohamed.khamis@glasgow.ac.uk}

\author{Eric Lecolinet}
\affiliation{%
  \institution{LTCI, Telecom Paris, IP Paris}
  \city{Palaiseau}
  \state{}
  \country{France}
}
\email{eric.lecolinet@telecom-paris.fr}

\author{Samuel Huron}
\affiliation{%
  \institution{CNRS i3 (UMR 9217)\\Telecom Paris, IP Paris}
  \city{Palaiseau}
  \state{}
  \country{France}
}
\email{samuel.huron@telecom-paris.fr}

\author{Jan Gugenheimer}
\affiliation{%
  \institution{LTCI, Telecom Paris, IP Paris}
  \city{Palaiseau}
  \state{}
  \country{France}
}
\email{jan.gugenheimer@telecom-paris.fr}

\renewcommand{\shortauthors}{Tseng, et al.}

\newcommand{\SteppingOn}{SteppingOn}
\newcommand{\HittingFace}{HittingFace}
\newcommand{\PerceptualMani}{perceptual manipulation} 
\newcommand{\VPPM}{VPPM} 

\begin{abstract}
``\textit{Virtual-Physical Perceptual Manipulations}'' (VPPMs) such as redirected walking and haptics expand the user's capacity to interact with Virtual Reality (VR) beyond what would ordinarily physically be possible. VPPMs leverage knowledge of the limits of human perception to effect changes in the user's physical movements, becoming able to (perceptibly and imperceptibly) nudge their physical actions to enhance interactivity in VR. We explore the risks posed by the malicious use of VPPMs. First, we define, conceptualize and demonstrate the existence of VPPMs. Next, using speculative design workshops, we explore and characterize the threats/risks posed, proposing mitigations and preventative recommendations against the malicious use of VPPMs. Finally, we implement two sample applications to demonstrate how existing VPPMs could be trivially subverted to create the potential for physical harm. This paper aims to raise awareness that the current way we apply and publish VPPMs can lead to malicious exploits of our perceptual vulnerabilities.
\end{abstract}

\ccsdesc[500]{Human-centered computing~Virtual reality}
\ccsdesc[500]{Human-centered computing~Human computer interaction (HCI)}

\keywords{virtual-physical perceptual manipulation, VPPM, physical harm, VR security}


\maketitle

\section{Introduction}
A particular direction of research at the intersection of Human-Computer Interaction (HCI) and Virtual Reality (VR) explores techniques that we define as \emph{Virtual-Physical Perceptual Manipulations} (VPPMs). \VPPM\ refers to Extended Reality (XR) driven exploits that \emph{alter the human multi-sensory perception of our physical actions and reactions to nudge the user's physical movements} \footnote{We consider this definition to be one of the outcomes of this paper. It was derived from insights from our workshop and  discussions among all the authors. We present this definition early on in our paper (rather than in the results section) to make it easier for the reader to follow.} (e.g., the position of body and hands). These techniques are often grounded in some threshold of the human perception (e.g., visual dominance \cite{posnerVisualDominanceInformationprocessing1976, vanbeersIntegrationProprioceptiveVisual1999}) and designed to overcome physical limitations of the current VR technology, enabling new types of interaction. Research focuses predominantly on positive intents, either discovering new {\VPPM}s \cite{lecuyerFeelingBumpsHoles2004, lecuyerPseudohapticFeedbackCan2000, burnsHandSlowerEye2005a} or presenting positive application scenarios for known {\VPPM}s. For example, redirection techniques are used to provide haptic feedback by changing the user's arm movement \cite{kohliRedirectedTouchingWarping2010, azmandianHapticRetargetingDynamic2016} or to enable a larger play area by steering the VR user's walking direction \cite{razzaqueRedirectedWalking2001, steinickeEstimationDetectionThresholds2010}. 

However, a {\VPPM} technique may vary in terms of prior consent and knowledge, and may also impact the user's ability to discern whether they are being manipulated. The user may be subjected to manipulation \emph{knowingly or unknowingly} and the manipulation \emph{may or may not be perceptible to the user}. Even if a user consents to being manipulated by {\VPPM}s, they might not be aware of the consequences of their physical actions because most {\VPPM}s are designed to be imperceptible to the user (i.e., below the perception threshold). Crucially, regardless of consent or knowledge, \emph{the intent behind a {\VPPM} is open to abuse} (e.g., disguising an attack as legitimate redirected walking) and may be opaque or covert to the user. This ambiguity, in terms of consent to, awareness of, knowledge of, and intent behind a given {\VPPM} is what gives rise to the significant potential for harm. Nothing is stopping malicious third parties from pursuing unknown, potentially harmful outcomes to the VR user using the perception thresholds published beforehand. The lack of a common definition has lead to a blind spot in research, where {\VPPM}s are proposed and published without due consideration as to their potential for harm.

In this paper, we focus on what is arguably the worst-case scenario --- imperceptible {\VPPM}s are applied to the user unknowingly, without consent. In particular, we focus on the potential for harm at an individual level, where one VR user's physical actions (i.e., body motions) are manipulated to a physically abusive end. We defined this physically abusive outcome as physical harm --- an action that causes hurt or damage relating to the VR user's body. The user is perceptually manipulated into physical action, and they perceive their agency while performing physical actions. Note that we focus on \emph{perceptual manipulation} as opposed to \emph{physical manipulation}. This means that approaches that physically manipulate the user through external devices such as Electrical Muscle Stimulation (EMS) \cite{lopesImpactoSimulatingPhysical2015} and exoskeletons are out of the scope of this paper. 
We exclude physical manipulations because these systems can physically direct or override the user's physical actions. Thus the user is implicitly aware of, having consented to this possibility through fitting these devices to their body. Whereas with {\VPPM}s, any physical actions are the result of a reaction to the presented perceptual stimuli, introducing the ambiguity around agency, intent, and consent of applying a \VPPM.
Based on this definition, we explore what potential physical harm could be provoked to the VR user by manipulating their physical actions through {\VPPM}s, and how malicious actors could potentially abuse {\VPPM}s to provoke physical harm.

The paper explores the risks posed by {\VPPM}s as follows. First, we demonstrate the potential threat of provoking physical harm using {\VPPM}s by presenting a threat model. A malicious actor wants to inflict physical harm on the VR user, and they can compromise the VR system by tricking the VR user into installing malware or a malicious app. 
Second, to be able to deeper understand these types of threats, we conduct a speculative design workshop using focus groups \cite{morganFocusGroupsQualitative1996, kruegerFocusGroupsPractical2015}. Because the physical harm exploited by {\VPPM}s is a novel phenomenon, our goal is to broadly explore the space and to promote discussion between participants. Using a design workshop \cite{jungk1987future, halskov06, ReflectionsDesignWorkshopRosner16} helps us to generate ideas and identify problems around the potential impact of the malicious use exploited by {\VPPM}s. We ran the workshop twice. The process of the workshops was video-recorded, transcribed, and coded using thematic analysis \cite{braunUsingThematicAnalysis2006}, unveiling 1) classifications of two main classes of attacks (\textit{puppetry} and \textit{mismatching}) using {\VPPM}s in VR and 2) the characterization of potential physical harm. Based on this classification of attacks, we present key publications in the HCI and VR community employing {\VPPM}s and note the lack of consideration given to malicious, subversive appropriation of this research. Finally, to demonstrate the process of subverting {\VPPM}s from existing publications in the field of HCI, we implement two sample applications (\SteppingOn\ and \HittingFace) based on two prior CHI publications, Haptic Retargerting \cite{azmandianHapticRetargetingDynamic2016} and Breaking the Tracking \cite{rietzlerBreakingTrackingEnabling2018}. We use both applications to demonstrate and reflect on our process, showing how concepts from \VPPM\ research could be trivially subverted to inflict malicious harm. We end this paper by discussing routes towards mitigating against, and preventing, malicious use of {\VPPM}s for practitioners and the research community.
This work has three contributions: 1) the definition of Virtual-Physical Perceptual Manipulation (\VPPM), classification of attacks, and characterization of physical harm that could be provoked by {\VPPM}s derived by two speculative design workshops (n=8); 2) two applications showing how we can trivially appropriate existing results of \VPPM\ research towards harmful intent; 3) mitigations and preventative recommendations for practitioners and the research community on how to deal with {\VPPM}s in the future.

\section{Threat Model}
In our threat model, an attacker wants to inflict physical harm on the VR user, and they can compromise the VR system. This can be done, for example, by tricking the VR user into installing malware or a malicious app. Similar to how smartphone spyware can use the affected smartphone sensors (e.g., as done in the Pegasus spyware\footnote{\url{https://en.wikipedia.org/wiki/Pegasus_(spyware)}}), the attacker can access information about the real-world environment around the VR user. This information can be extracted from tracking devices like the front-facing headset camera(s) used for inside-out tracking. The attacker can also exploit the sensors inside the VR headset and the controllers to understand the user's movement in real-time, or access all the standard APIs that are normally available to VR applications. A sample scenario is that a user is tricked into installing a malicious VR app that contains {\VPPM}s that do not specify their intent and are disguised as a part of the application. The user is thus presented with a VR setup that manipulates them imperceptibly (e.g., walking, reaching objects). Because the attacker has access to information about the user's motion and the safety boundaries (e.g., Oculus Guardian), the attacker can inflict physical harm on the user through the setup. Examples of harm include tripping, hitting a wall, holding something dangerous, or walking into a dangerous area in the context of accidents \cite{daoBadBreakdownsUseful2021} and bystander abuse \cite{ismar2021ohagan}. Such harms may have significant implications on the user including even death \cite{wildeManDiesVR2017}.

\section{Related Work}
Our work builds on prior research in presence, perceptual manipulations, ethics and security in VR.

\subsection{VR Technologies and Experiences}
VR technologies track the user's physical actions in a 3-D space using Head-Mounted Displays (HMDs) and controllers, providing stimuli (e.g., visual, auditory, haptic) to enable embodied interactions. Through these technologies, VR elicits strong immersive experiences that allow the user to have a subjective feeling of being present in a virtual environment and act realistically, despite the VR user consciously knowing that the virtual environment does not physically exist. The sense of being in a virtual environment is called \emph{presence} \cite{sheridanMusingsTelepresenceVirtual1992, slaterFrameworkImmersiveVirtual1997, draperTelepresence1998} in VR. For example, participants tend to take a longer path on the simulated ground rather than walking over a virtual pit \cite{meehanPhysiologicalMeasuresPresence2002}. VR users can also feel that the events happening in the virtual environment are real (e.g., plausibility illusion \cite{sanchez-vivesPresenceConsciousnessVirtual2005, slaterPlaceIllusionPlausibility2009}) and that the virtual body parts or even a full-body avatar have become a part of their own (e.g., embodiment illusion \cite{spanlangHowBuildEmbodiment2014}). These illusory states in VR are the outcomes of our perception and do not directly affect our higher cognitive functions \cite{gonzalez-francoModelIllusionsVirtual2017}. Enhancing the immersive experience and presence in VR becomes a common goal for designing new VR interaction or locomotion techniques. The existence of these illusions and the fact that they are working so well, is one of the main reasons why {\VPPM}s can be applied so effortless to a variety of application scenarios. 

\subsection{Perceptual Manipulations in VR}
VR is an excellent platform for applying perceptual manipulation. While {\VPPM}s can apply across the reality-virtuality continuum, we focus on VR because of its greater capacity for inducing an illusion of non-mediation. The simulated content occupies the VR user's visual sensory input, and VR HMDs block the user's view of the outside world to enhance immersion. These features allow designers to make use of the visual dominance \cite{posnerVisualDominanceInformationprocessing1976, vanbeersIntegrationProprioceptiveVisual1999} and the unawareness of sensory discrepancy \cite{warrenVisualproprioceptiveInteractionLarge1971, vanbeersWhenFeelingMore2002}.

Research in HCI and VR develops techniques to manipulate the mapping between virtual and physical environments. Most of the time, they are below the human perception threshold, making them imperceptible. Previous research found that one can induce the pseudo-haptic feedback by controlling the visual input \cite{lecuyerPseudohapticFeedbackCan2000, lecuyerFeelingBumpsHoles2004} and that VR users are less sensitive to the visual-proprioceptive conflict \cite{burnsHandSlowerEye2005a}. Although there is a difference between the virtual and physical environment, our perceptual system interprets the sensory information from VR, and the brain-body system reacts immediately to perform the physical actions \cite{gonzalez-francoModelIllusionsVirtual2017}.

Practitioners and researchers then start ``hacking'' human perception to overcome several limitations in current VR systems (e.g., limited tracked space, lack of haptic feedback). A popular example of such technique is redirected walking \cite{razzaqueRedirectedWalking2001, steinickeEstimationDetectionThresholds2010, sunVirtualRealityInfinite2018}, steering the VR user's physical walking path by interactively and imperceptibly rotating the virtual scene. One can use slow-speed translation/rotation gain below the user's perception threshold or manipulate the stereo image in a see-through HMD \cite{ishiiOpticalMarionetteGraphical2016} to achieve the effect. Redirected touching \cite{kohliRedirectedTouchingWarping2010} and redirected haptics \cite{azmandianHapticRetargetingDynamic2016, chengSparseHapticProxy2017} re-purpose the VR user's hand to a passive haptic prop by manipulating the visual of the user's arm or the virtual scene. These manipulations can also be applied to reduce physical movements and fatigue by improving ergonomics in VR \cite{montanomurilloErgOErgonomicOptimization2017}, changing VR user's posture unobtrusively \cite{shinBodyFollowsEye2020}, and inducing a sensation of weight \cite{rietzlerBreakingTrackingEnabling2018, samadPseudoHapticWeightChanging2019}.

While the aforementioned are applications of {\VPPM}s that had positive intents, {\VPPM}s can also be exploited maliciously to provoke harm on the VR user. The adversaries in that case can be VR developers who intentionally (or unintentionally) manipulate the user's perception in a way that has harmful consequences. Our method is to articulate how {\VPPM}s in VR can be --- and likely will be --- abused in the future.

\subsection{The Potential Harm and Attacks in VR}
VR induces strong sensory feedback on our perception. Previous work discussed the ethical implications of conducting VR research \cite{behrPracticalConsiderationsEthical2005, madaryRealVirtualityCode2016} and of realism in VR and Augmented Reality (AR) \cite{slaterEthicsRealismVirtual2020}. In our work, we focus on uses of VR that are highly persuasive for benefits (e.g., training), but could also be used for malicious purposes. 
An example would be to incite a VR user to do something they would not normally do, which in turn leads to harming the VR user. 

Through the \VPPM\ techniques, one can change the VR user's perception of their physical actions. Current VR applications are dominantly achieved through embodied motions for enhancing presence \cite{usohWalkingWalkinginplaceFlying1999}. Compared to interaction with desktop or mobile devices, VR involves a larger-scale of 3-D space, which means that the VR user is more likely to encounter physical harm caused by their actions. An example is to elicit the user to sit on a virtual chair that does not have a counterpart in the real world. More examples, like colliding or hitting real-world objects and falling over, have been identified in a recent work on common VR fails that happen to users at home \cite{daoBadBreakdownsUseful2021}.

Recently, security researchers started to explore the potential for immersive VR attacks. Casey \textit{et al.} \cite{caseyImmersiveVirtualReality2019}, presented a software vulnerability and were able to manipulate the visuals of the safety guardians of an HTC VIVE. Using this, the authors identified what they called the ``Human Joystick Attack'', which allows directing an immersed user's physical movement to a location without the user's knowledge. This attack falls under one of the five categories we identify in our classification of malicious \VPPM\ use. Our work extends this previous research by understanding the larger class of attacks that could be possible using {\VPPM}s. While the security community started to explore potential vulnerabilities in XR technologies, the current main focus is on finding and closing new factors of attack on the software and hardware \cite{odeleye2021detecting, 9319051}. However, in this work we are not focusing on the technical weak spots but are actually exploring human weak spots. We argue that the HCI community is at the perfect intersection of computer science, psychology, cognitive science and design, to combine knowledge from those fields who are mostly publishing {\VPPM}s.

\section{Method: Speculative Design Workshop}

Because we want to understand what malicious exploits of \VPPM\ might look like in the future, we refer to methods such as speculative design \cite{augerSpeculativeDesignCrafting2013} and design fiction \cite{markussenPoeticsDesignFiction2013}. These approaches allow us to both critique current practices and reflect on future technologies and their ethical implications. We broadly explore this space through a speculative design workshop using focus groups \cite{morganFocusGroupsQualitative1996, kruegerFocusGroupsPractical2015} with researchers and designers. Participants had to a) brainstorm scenarios in which {\VPPM}s can be used to induce physical harm to the VR user; b) identify one (or more) dimension upon which the scenarios from the brainstorming can be contextualized (e.g., the severity of physical harm); and c) rate the relevance of each dimension for studying and preventing future physical harms caused by {\VPPM}s in VR. 

\subsection{Participants}
We used snowball sampling and reached out to people from the mailing list. Eight participants (age: $M=28.3, SD=2.1$) were recruited (Table \ref{tab:workshop-participants}). All researchers worked on VR/XR topics, publishing peer-reviewed papers in top-tier conferences like CHI and UIST. To get a more diverse group of people, ideas and perspectives, we additionally recruited participants that identified their work to be dominantly on design rather than on research or development. We argue there is a benefit in having a range of expertise as experts alone may be overly constrained in their thinking based on their knowledge of technical constraints or prior research \cite{crossExpertiseDesignOverview2004}. Therefore it was important to have that blend of expert and non-expert/familiar participants. We also want to clarify that none of our participants were novices in the field of XR. Most of our participants rated their VR expertise to be at least average and mostly above average and expert. Overall, they had at least average and above-average experience with VR (5-point Likert scale, $M=3.75, SD=1.03$). We ran the speculative design workshop twice with four participants each time.

\begin{table}[t]
  \caption{The background of participants. We asked participants to self-describe their profession and VR expertise.}
  \Description[Table 1 indicates the background of participants.]{Table 1 indicates the background of participants, including participant ID, gender, profession, and self-described VR expertise.}
  \label{tab:workshop-participants}
  \begin{tabular}{ccll}
    \toprule
    ID & Gender & Profession & VR Expertise\\
    \midrule
    W1P1 & F & HCI researcher      & expert\\
    W1P2 & M & HCI researcher      & above average\\
    W1P3 & M & HCI researcher      & above average\\
    W1P4 & M & XR/HCI designer     & expert\\
    W2P5 & M & curator/designer    & below average\\
    W2P6 & F & design researcher   & average\\
    W2P7 & F & HCI researcher      & above average\\
    W2P8 & F & graphic/interaction designer  & average\\
    \bottomrule
  \end{tabular}
\end{table}

The goal was to explore scenarios using {\VPPM}s to provoke physical harm. With our introduction in the workshop, participants could design a malicious scenario using \VPPM. P1, P2, P3, and P7 had a Computer Science background and worked on HCI and VR/XR research. P4 worked as an XR/HCI designer from the industry, who develops VR training platforms for surgeries. P5, P6, and P8 were designers who have a design background working in design research. A designer could think about diverse contexts and consequences of the abusive scenario, and a researcher could deep dive into the technical details if they consider it is necessary. Both workshops included researchers and designers, the first one (W1) was more researcher-focused, and the second one (W2) was more designer-focused. This setup allowed each workshop to enable discussions with different perspectives and elicit valid outcomes.

\subsection{Procedure}

\begin{figure}[t]
\begin{center}
  \begin{tabular}{@{\hspace{0.1cm}}c}
		\includegraphics[width=\linewidth]{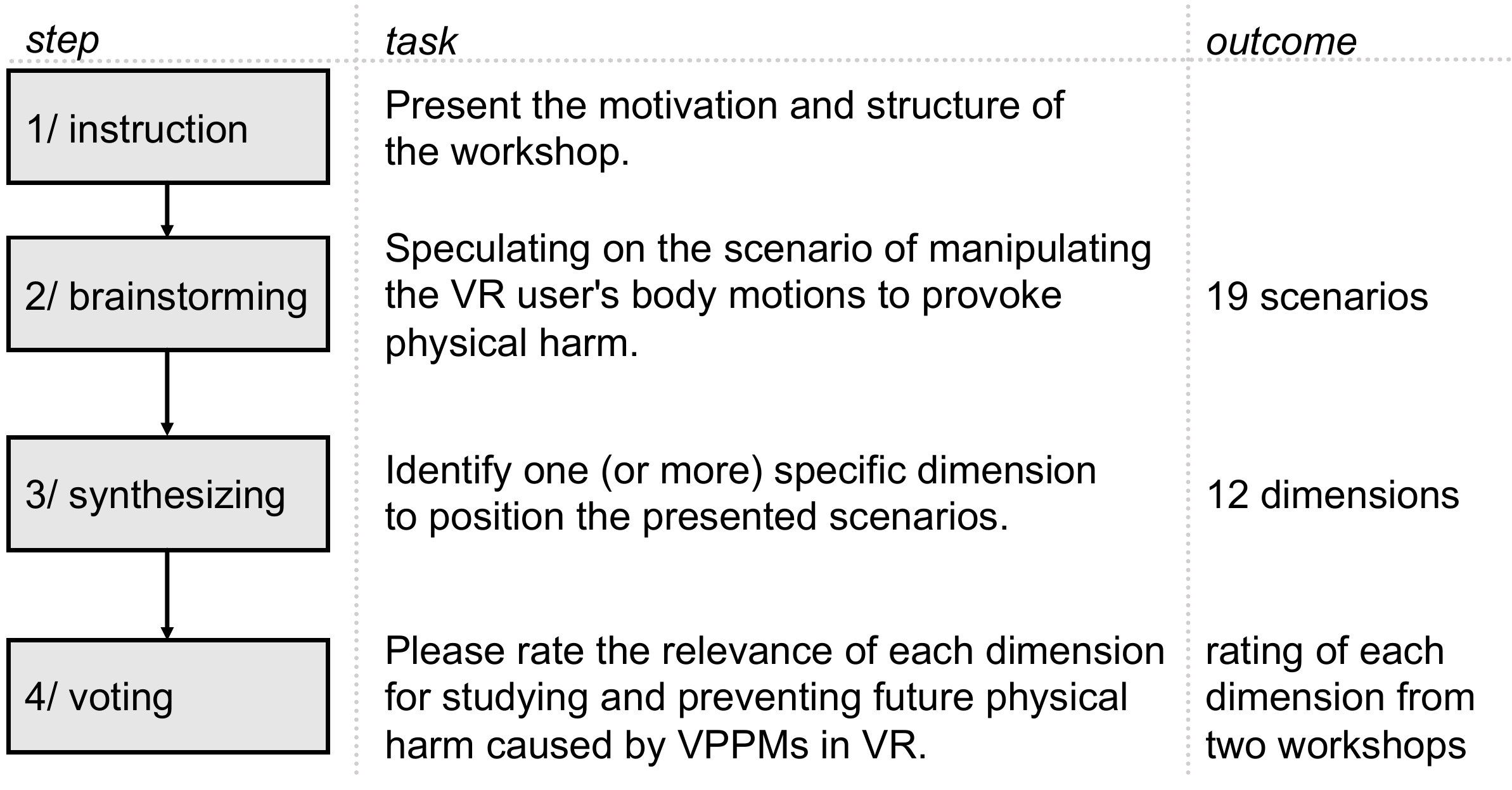}
  \end{tabular}
    \caption{The steps, tasks and outcomes of the speculative design workshop.}
    \label{fig:workshop-structure}
    \Description[Figure 1 shows the process, task, and outcome of the workshop.]{Figure 1 shows four steps with their tasks and outcomes in the speculative design workshop. We ran two workshop and obtained overall 19 scenarios and 12 dimensions.}
\end{center}
\end{figure}

Figure \ref{fig:workshop-structure} shows the structure of our speculative design workshop. The workshop consisted of four steps: instruction, brainstorming, synthesizing, and voting.
In the instruction step, we first introduced the VPPMs in VR by presenting examples in HCI and VR research, such as Haptic Retargeting \cite{azmandianHapticRetargetingDynamic2016}, Body Follows Eye \cite{shinBodyFollowsEye2020}, and redirected walking \cite{ishiiOpticalMarionetteGraphical2016, sunVirtualRealityInfinite2018}. Next, we presented our goal --- speculate on the potentially abusive {\VPPM}s that could manipulate the VR user's body motions to induce physical harm. This part took 15 minutes to complete.

In the brainstorming step, we presented the following assumption: ``\textit{In 10 to 20 years, VR technology has full body tracking and understands the physiological states of the VR user. People can use VR in open space, and VR application becomes more than gaming and lab experiments. {\VPPM}s are able to manipulate whole-body motions and are imperceptible to the VR user.}'' Based on this assumption, we introduced the task:
\begin{quote}
    \textbf{Brainstorming Task:} Speculate on a scenario manipulating the VR user's body motions to provoke physical harm.
\end{quote}
Participants had to describe how they use {\VPPM}s to elicit physical actions that provoke physical harm. One restriction in the brainstorming was that the VR user has to perceive agency on their physical actions. We do not consider body motions created by an external device (e.g., EMS or exoskeleton) as {\VPPM}s because the VR user knows the motion is done by the system. Participants had 10 minutes time to brainstorm as many scenarios as they could individually. Afterwards, each participant presented their ideas and discussed it with the other participants (15 minutes).

After participants presented their scenarios, we continued with the third step: 
\begin{quote}
    \textbf{Synthesizing Task:}
    Identify one (or more) specific dimension to position the presented scenarios.
\end{quote}
The goal of synthesizing was to understand the potential harm in more detail that could happen using {\VPPM}s. We asked participants to find one or more specific dimension that can be used to position all the presented scenarios on (including the ones from other participants). The goal was to find terms and variables that are helpful to understand the potential harm. One example could be to use ``amount of pain'' as the variable and position scenarios that create little pain further on the left than scenarios that create more pain. Participants created dimensions individually for 10 minutes and took turns to present their outcomes altogether for 10 minutes.

Finally, in the voting step, we asked participants to rate the relevance of each dimension created in the synthesizing step:
\begin{quote}
    \textbf{Voting Task:} Please rate the relevance of each dimension for studying and preventing future physical harm caused by {\VPPM}s in VR.
\end{quote}
The rating was a 5-point Likert scale ranging from strongly irrelevant to strongly relevant. Note that the two workshops had different scenarios and dimensions. The W1 participants rated the dimensions created in W1 and the same for W2. Here we were interested in the consensus of the participants in each workshop. This part took five minutes to complete.

All participants engaged in the discussion during both the brainstorming and synthesizing steps. The discussion allowed participants to collaborate in groups to discuss the scenarios and dimensions they created. Therefore participants worked together to create the outcome. All of them contributed to the question about the potential malicious use of {\VPPM}s (scenarios in the brainstorming step) and the range of the presented scenarios (dimensions in the synthesizing step). Participants worked on miro\footnote{\url{https://miro.com/} (Last access: 9th Apr. 2021)} remotely, and the both workshops lasted two hours. We recorded the brainstorming and the synthesizing steps.

\section{Workshop Results}
In this section, we first introduce the analysis of the results from our speculative design workshops. Next, the collected data and extracted results (e.g., including the classification of attacks and the characterization of physical harm) are presented. Finally, we summarize the observations from the workshop.

\subsection{Data Analysis}
Figure \ref{fig:workshop-structure} (the right column) shows the outcome of each step of the workshop. Participants from the two workshops created 19 scenarios and 12 dimensions. The video footage of the workshops was transcribed and anonymized. The transcripts and scenarios were then iterated and coded by three authors in joint sessions. Participants did not take part in the analysis. We applied thematic analysis \cite{braunUsingThematicAnalysis2006} to investigate the underlying themes of the transcribed data. The coding was always done together in nine sessions, each of which took on average two hours. Several sessions were re-watched during the coding sessions to arrive at a consistent interpretation consisting of categories and general themes. Conflicts were resolved by discussing each individual coding.

\subsection{Scenarios}
\begin{figure*}[t]
\begin{center}
  \begin{tabular}{@{\hspace{0.1cm}}c}
		\includegraphics[width=\linewidth]{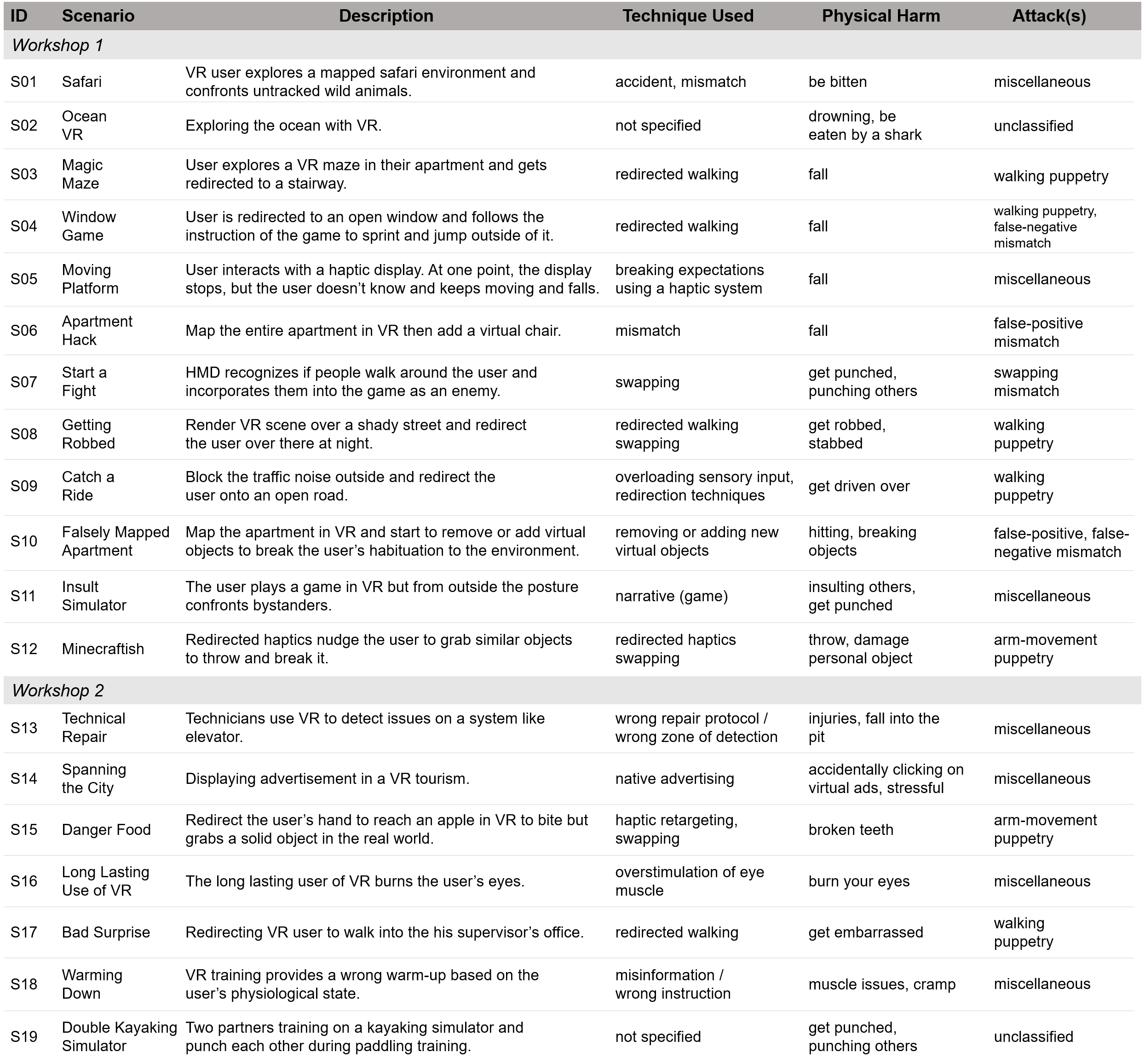}
  \end{tabular}
  \caption{An overview of 19 scenarios we collected, including the name, description, technique used, physical harm, and classified attack of each scenario.}
  \Description[Figure 2 shows an overview of the 19 scenarios elicited from the workshop.]{Figure 2 shows the name and description of 19 scenarios. Technique used and physical harm are the code we created in the thematic analysis. The attack column shows the classified attack.}
  \label{fig:scenario-all}
\end{center}
\end{figure*}

Figure \ref{fig:scenario-all} presents 19 scenarios: names, descriptions, techniques used to induce them, and the potential physical harm caused in them. \textit{Technique Used} and \textit{Physical Harm} are the codes identified in the thematic analysis. Several scenarios apply redirection techniques to affect the user's physical movements or actions and bring them to harmful consequences: \textit{Magic Maze}, \textit{Window Game}, \textit{Bad Surprise}, \textit{Minecraftish}, \textit{Danger Food}, \textit{Getting Robbed}, and \textit{Catch a Ride}. Three scenarios try to break the habituation and trust of using a system to provoke physical harm (\textit{Apartment Hack}, \textit{Falsely Mapped Apartment}, \textit{Moving Platform}). Some scenarios occlude the physical world with virtual content so the VR user is unaware of the physical harm: \textit{Getting Robbed}, \textit{Start a Fight}, \textit{Safari}, \textit{Ocean VR}. \textit{Insult simulator} uses game instructions to make inappropriate gestures to insult the bystanders. The rest of the five scenarios are not directly associated with {\VPPM}s. \textit{Technical Repair} and \textit{Warming Down} provide false information to induce harm. \textit{Spanning the City} is a scenario about advertisement in VR. \textit{Double Kayaking Simulator} does not specify the technique, and \textit{Long Lasting Use of VR} is about overusing VR. The description of seven selected scenarios are presented in Appendix \ref{apx:scenarios} as a representation of those using similar techniques.

\subsection{Classification of Attacks}
To find some commonalities and a potential classification of attacks, we applied open and axial coding on the \textit{Technique Used} that was presented with each scenario (the label that described how participants wanted to achieve the effect). The identified codes were shown in Figure \ref{fig:scenario-all}, the \textit{Attack(s)} column. We identified two main classes of attacks:  \emph{puppetry} attacks and \emph{mismatching} attacks. For the scenarios that did not reach a greater theme, we coded them as \emph{miscellaneous}. They provided different insights like accidents (S05, S13) or social interaction (S11). Two scenarios (S02 and S19) were coded as unclassified because they were too specific and missed the technical detail. In the following, we focus on the definition of \emph{puppetry} and \emph{mismatching} attack and how they are integrated into the scenarios.

\begin{figure*}[t]
\begin{center}
  \begin{tabular}{@{\hspace{0.1cm}}c}
		\includegraphics[width=\linewidth]{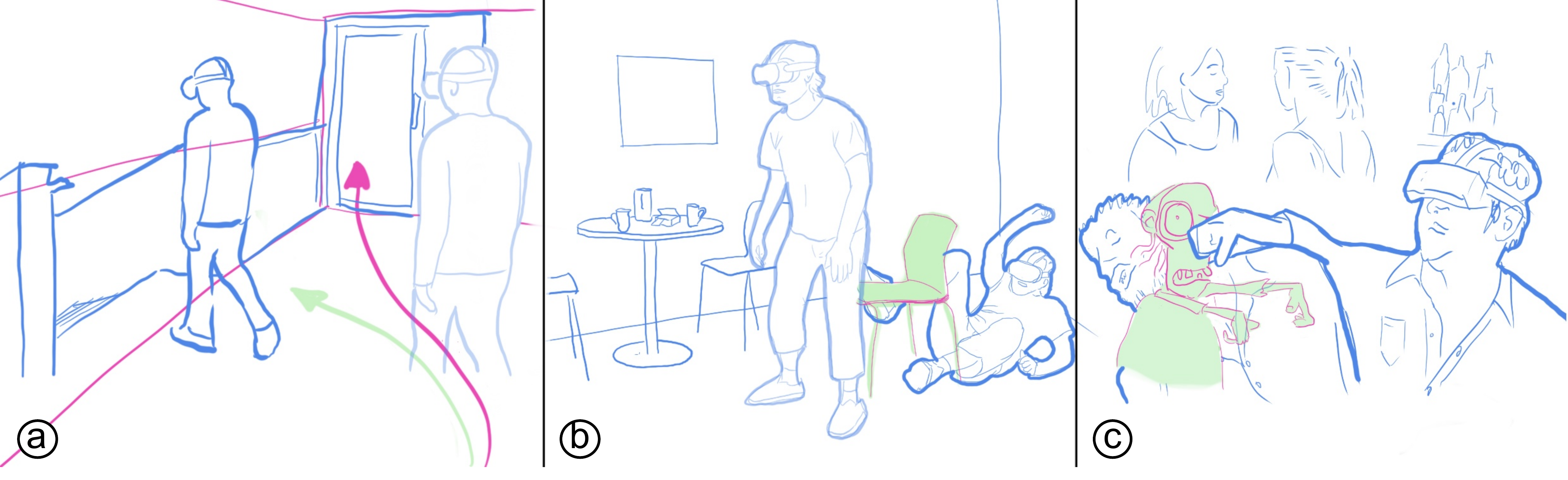}
  \end{tabular}
  \caption{We illustrate the attacks by showing the sketching of three selected scenarios. For the color code, the blue outline represents the physical world, pink stands for the virtual content, and green shows how attack works. (a) The VR user is in \textit{Magic Maze} scenario and thinks they walk along the direction to the physical door (the purple arrow). Malicious actors apply the \emph{walking puppetry} attack to steer the VR user's walking direction and make them fall off a stairway. (b) The VR user locates in a fully-mapped apartment (\textit{Falsely Mapped Apartment}). Malicious actors apply the \emph{false-positive mismatching} attack to introduce a virtual chair. The user assumes the virtual chair is fully-mapped. So they sit on the chair, but end up falling on the floor. (c) A VR user is playing a zombie game where they have to fight with zombies using bare hands. Malicious actors use the \emph{swapping mismatching} attack render the virtual zombie over a bystander and makes them start a fight. }
  \Description[Figure 3 shows three selected scenarios and illustrates the walking puppetry, false-positive, and swapping attacks.]{Figure 3 demonstrates the attacks by showing the sketching of three selected scenarios. Figure 3a illustrate the Magic Maze scenario where the VR user walks along the arrow pointing towards the door. Malicious actors apply walking puppetry attack to steer the VR user's walking direction towards the stair and make them fall off the stairway. Figure 3b shows the scenario Falsely Mapped Apartment. The VR user habituates to a fully mapped room. Malicious actors apply false-positive mismatching attack to introduce a virtual chair, and the user sits on the virtual chair, ending up falling on the floor. Figure 3c shows the scenario start a fight. A VR user plays a zombie game where they have to fight with zombies using bare hands. Malicious actors use the swapping mismatching attack render the virtual zombie over a bystander and makes them start a fight.}
  \label{fig:attacks}
\end{center}
\end{figure*}

\subsubsection{\textbf{Puppetry Attacks}}
These attacks control \emph{physical actions of different body parts of an immersed user}. We argue that {\VPPM}s allow controlling different body parts precisely as the technology and research progress. Therefore we use the term ``puppetry'' to represent the potential impact that this attack could happen on different levels of body parts in the future.


\paragraph{Walking Puppetry Attack}
By applying redirected walking VPPMs, the malicious actor can steer the VR user's walking direction (Figure \ref{fig:attacks}a). The walking puppetry attack was mentioned in several scenarios, including \textit{Magic Maze} (S03), \textit{Window Game} (S04), \textit{Getting Robbed} (S08), \textit{Catch a Ride} (S09), and \textit{Bad Surprise} (S17). Participants applied this attack to make a VR user go to a location for provoking potential physical harm (e.g., falling, going to a dangerous area).

\paragraph{Arm-Movement Puppetry Attack}
The arm-movement puppetry attack controls the physical actions of the VR user's arm. Redirected haptic techniques \cite{kohliRedirectedTouchingWarping2010, azmandianHapticRetargetingDynamic2016} are the underlying {\VPPM}s. By applying this attack, one can direct a VR user's hand to interact and break the user's property (\textit{Minecraftish}, S12) or to reach a physical object that could be harmful during interaction (\textit{Danger Food}, S15). 

\subsubsection{\textbf{Mismatching Attacks}}
\emph{Mismatching} attacks are manipulations in which the adversary exploits a difference of information between a virtual object and its physical counterpart to elicit misinterpretation for the VR user. Here the environment for a VR user is true-positive, where each virtual object has a one-to-one representation in the real world.
 
\paragraph{False-Positive Attack}
In Figure \ref{fig:attacks}b, the false-positive attack creates virtual content that has no physical counterpart (e.g., a virtual chair) in a true-positive environment. The VR user habituates this one-to-one mapping environment therefore they believe the false-positive chair exists in the room. Interacting with these content could lead to physical harm (e.g., sitting on a virtual chair and falling on the floor).
Scenarios using the false-positive attack are \textit{Falsely Mapped Apartment} (S10) and \textit{Apartment Hack} (S06). They require a perfectly mapped environment and a certain degree of trust from the user towards the VR environment. This trust, most of the time, builds upon how much a VR user is accustomed to the environment or interaction. In fact, the habituation can be achieved through repeating a single task. Once the user is used to the task and starts performing it without conscious attention, a false-positive attack becomes dangerous and impactful. 

\paragraph{False-Negative Attack}
In this attack, the malicious actor deliberately hides the information from the physical environment. Therefore, the VR user is unaware of incoming dangers. For example, overriding traffic noise (\textit{Catch a Ride}, S09) makes the user unaware of approaching vehicles, which in turn makes them vulnerable. In \textit{Falsely Mapped Apartment} (S10), malicious actors provoke collision with the environment by removing an virtual object from a fully-mapped apartment.
The false-negative attack could happen when using VR in an open space because the system needs to constantly detect the surroundings. If the attack hides or disguises a physical object (e.g., hiding an opened window), the attacker could make the VR user even fall or jump out of this window (\textit{Window Game}, S04). Which could even lead to a fatal outcome.

\paragraph{Swapping Attack}
The swapping attack happens in the True-Positive situation where each virtual object maps to a physical object. However, the application renders a different virtual image that does not represent the identity of the physical object. Therefore, the VR user believes they are interacting with the virtual one but inadvertently cause physical harm to themselves or to others.
In \textit{Start a Fight} (S07), the bystanders were rendered as the enemy avatars in a fighting VR game, which resulted in the VR user attacking bystanders (Figure \ref{fig:attacks}c). 


\subsubsection{Reflection on VPPM research and Potential Attacks}
While some of the presented scenarios may stretch the imagination, we want to emphasize that for the most part these scenarios already exist in some prior work that started to work towards the potential \VPPM\ and potential abuse. To demonstrate this we selected for every type of attack a few example publications from the field of HCI. We selected publications that were either working towards a \VPPM, or presented a new application of {\VPPM}s which could be used to reproduce the work.
We want to emphasize that this is by far not an exhaustive list but should only work as en example. 

For \emph{puppetry} attacks, we select seven papers 
\cite{razzaqueRedirectedWalking2001, steinickeEstimationDetectionThresholds2010, ishiiOpticalMarionetteGraphical2016, langbehnBendingCurveSensitivity2017, langbehnBlinkEyeLeveraging2018, sunVirtualRealityInfinite2018, rietzlerRethinkingRedirectedWalking2018} in which the walking attack is possible, and four \cite{kohliRedirectedTouchingWarping2010, azmandianHapticRetargetingDynamic2016, rietzlerBreakingTrackingEnabling2018, samadPseudoHapticWeightChanging2019} in which the arm-movement attack is possible. These publications mainly investigated redirection techniques and are often published at AR/VR conferences such as ISMAR, IEEE VR and UIST. In these papers, there are few hardware requirements (although some do need eye-tracking) and the implementations are described in detail. The thresholds of applying {\VPPM}s are also provided in these publications.
For \emph{mismatching} attacks, we selected the following publications (false-positive: \cite{chengVRoamerGeneratingOnTheFly2019, yangDreamWalkerSubstitutingRealWorld2019, remixing}, false-negative: \cite{hartmannRealityCheckBlendingVirtual2019a,remixing}, swapping: \cite{simeoneSubstitutionalRealityUsing2015, hettiarachchiAnnexingRealityEnabling2016, shapiraRealitySkinsCreating2016}). Among the selected publications, only Optical Marionette \cite{ishiiOpticalMarionetteGraphical2016} mentioned the safety concern of manipulating the user's walking in the real world. There is a lack of consideration given to malicious, subversive appropriation of {\VPPM} research.



\subsection{Characterizing Physical Harm}
In the synthesizing step, we asked participants to identify one (or more) specific dimension to position the presented scenarios.
We report two dimensions (\textit{severity of physical harm} and \textit{perceived agency}) that received the highest score in the voting step of each workshop. Note that each workshop had a different output of scenarios and dimensions. Therefore, the consensus of the voting is within each workshop. Finally, we report on our last analysis of characterization of the physical harm we found in the workshop.

\begin{figure}[t]
\begin{center}
  \begin{tabular}{@{\hspace{0.1cm}}c}
		\includegraphics[width=\linewidth]{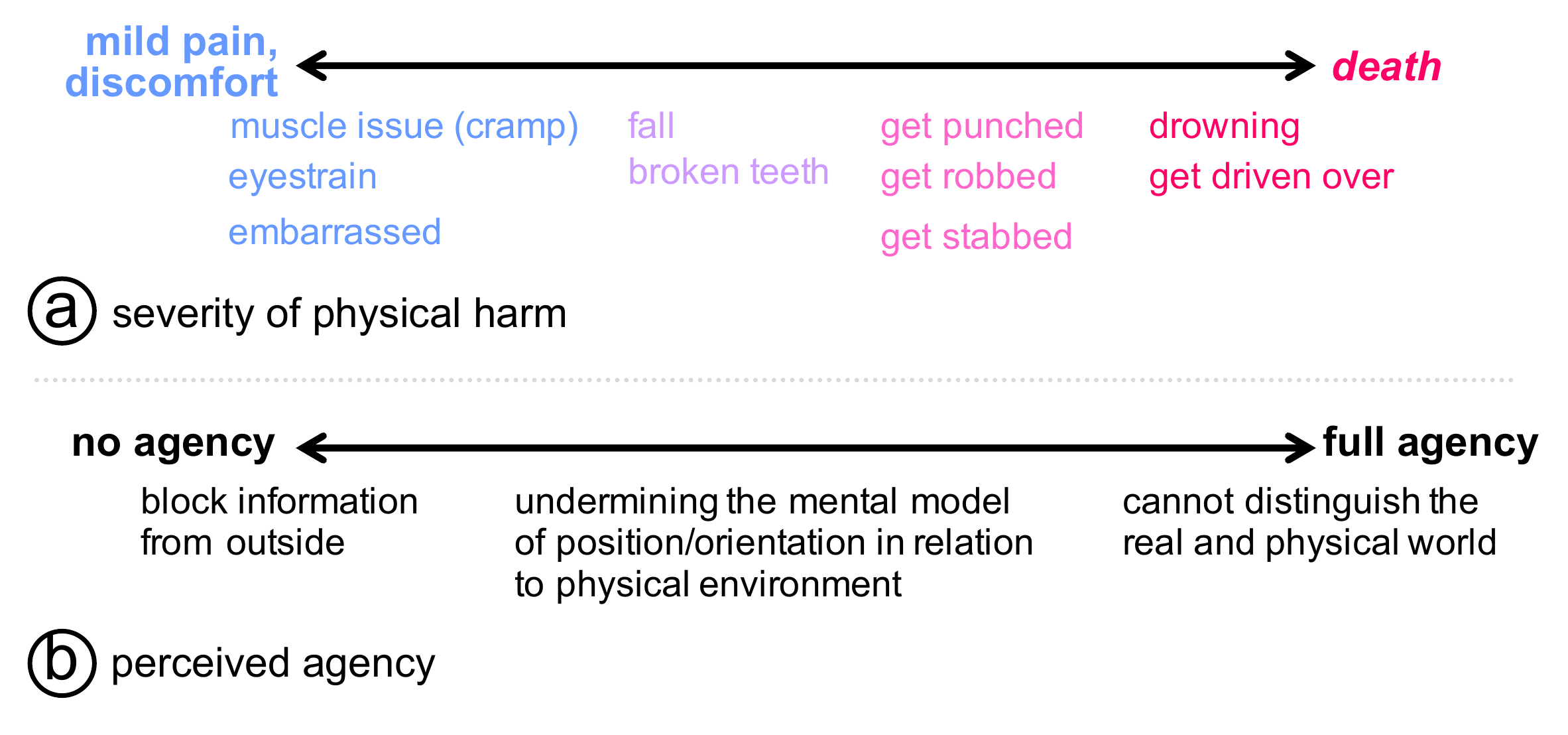}
  \end{tabular}
  \caption{We reported two dimensions selected from the synthesizing step. (a) \textit{Severity of the Physical Harm} shows how bad physical harm can be and can a VR user recover from the given harm. This dimension varies from mild pain and discomfort (e.g., eyestrain, cramp) to the extreme case (e.g., drowning, get driven over). (b) The \textit{Perceived Agency} indicates to what degree a VR user considers that the physical harm (or consequence) is caused by themselves.}
  \Description[Figure 4 shows two dimensions created in the synthesizing step, severity of physical harm and perceived agency.]{Figure 4a shows the dimension, severity of the physical harm, varies from little pain and discomfort (e.g., eyestrain, cramp) to the extreme harm (e.g., drowning, get driven over). Figure 4b illustrates the dimension, perceived agency. This dimension shows to what degree a VR user believe the physical harm or the consequence of the interaction is caused by themselves.}
  \label{fig:scenario-harm}
\end{center}
\end{figure}

\paragraph{\textbf{Severity of the Physical Harm}}
Overall, 16 instances of physical harm were mentioned, in which falling and punching each appeared four times. The severity of the physical harm is the most reported dimension in the synthesizing step (6 out of 12 dimensions). We interpret severity as how bad the physical harm can be on a VR user and can the VR user recover from the given physical harm. Figure \ref{fig:scenario-harm}a, from the left, the physical harm is a low, brief moment of discomfort (e.g., eyestrain, falling, punches). From the right, the physical harm becomes more unrecoverable (e.g., broken teeth, get driven over), and the extreme form of severity is death. 

\paragraph{\textbf{Perceived Agency}}
The \textit{Perceived Agency} (D2) is one of the dimensions reported by participants in the synthesizing step (Figure \ref{fig:scenario-harm}b). The \textit{Perceived Agency} is to what degree VR users consider the harm is caused by themselves. No agency means the VR user interprets the system (or application) caused the physical harm. For instance, If a user finds out the system blocks all the auditory information from outside but does not maneuver this setting, he perceives no agency in this case. On the other hand, full agency means VR users perceive the harmful consequence is done by themselves. The implication from the perceived agency dimension is whether a VR user falls into the same trick again. Nevertheless, because the {\VPPM}'s manipulation may or may not be perceptible, a malicious exploit of \VPPM\ can hide their maneuver on the user and make them blame themselves.



\paragraph{\textbf{The Origin of Harm Created}}
Similar to how we coded the \textit{Technique Used} to classify types of attacks, we now coded  \textit{Physical Harm} to find a classification of harm. 

\begin{figure}[t]
\begin{center}
  \begin{tabular}{@{\hspace{0.1cm}}c}
		\includegraphics[width=\linewidth]{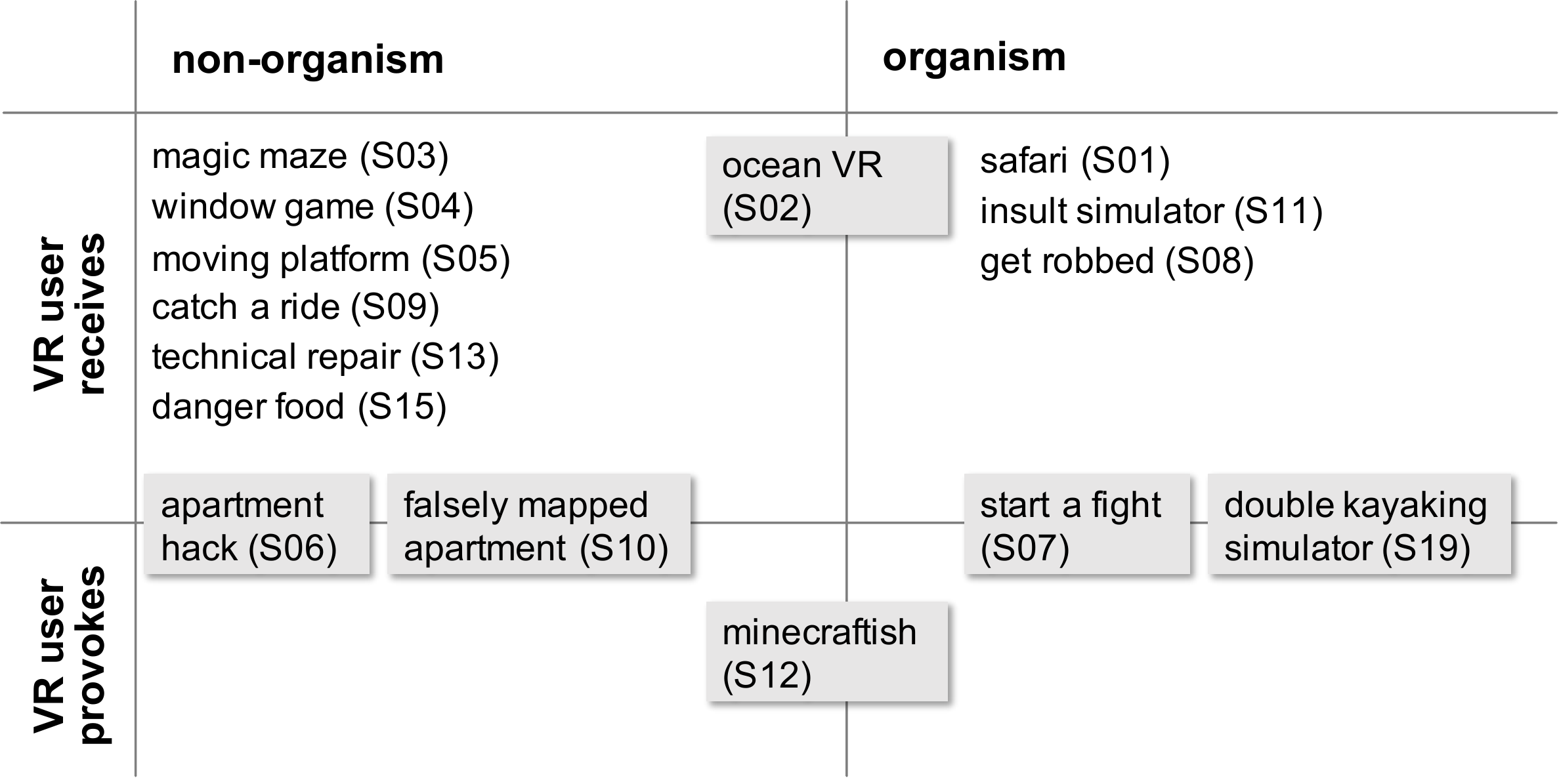}
  \end{tabular}
  \caption{The matrix categorizes the physical harm by 1) VR user provokes/receives the physical harm and 2) whether the other party is an organism or non-organism. Scenarios with the gray background fit into two quadrants.}
  \Description[Figure 5 describes a categorization of physical harm, the origin of harm created.]{Figure 5 describes a categorization of physical harm, the origin of harm created. We categorize the physical harm by whether a VR user provokes/receives the physical harm and whether the other party is an organism or non-organism. Scenarios with the gray background fit into two quadrants.}
  \label{fig:harm-whotowhom}
\end{center}
\end{figure}

We find in the dimension of severity that physical harm can be caused by the user (e.g., fall down into stairway) or by others (e.g., someone punches the VR user). This was also mentioned in the \textit{origin of physical harm done} (D6) in the synthesizing step. Therein, participants described who committed the physical harm in each scenarios. We extend this concept in our coding process and present a 2 $\times$ 2 matrix (Figure \ref{fig:harm-whotowhom}) to categorize physical harm by 1) VR user provokes/receives the harm and 2) is the other party an organism or non-organism.

Scenarios with the gray background fit into two quadrants at the same time. Because our task focused on inducing physical harm to the single VR user, most scenarios locate in the quadrant of receiving harm from non-organism (e.g., falling down a stairway, get driven over a car). Although we asked participants to create physical harm related to the VR user's body, damage to non-organism still came up during the workshop. For example, hitting furniture, breaking personal property by throwing them. We categorize these property damages into VR user provokes harm to non-organism. The VR user also provokes physical harm to organism (e.g., punching bystanders in \textit{Start a Fight}, throwing a pet in \textit{Minecraftish}). Finally, the VR user can also get hurt from organism. An example would be get stabbed in \textit{Get Robbed} or bitten by wild animals in \textit{Safari}. This matrix indicating physical harm could be extended to more than one VR user in the future.

\subsection{Observations from the Workshop}
Among all the scenarios, seven (7/19=37\%) of them applied \emph{puppetry} attacks as a part of the technique used to provoke physical harm to the VR user. The \emph{puppetry} attack was several times combined with mismatching attacks (e.g., \textit{Catch A Ride}: false-negative + walking puppetry) and easier to apply and deploy in VR applications. Therefore, they have the potential to become of the first archetypes of malicious attacks using {\VPPM}s.


\paragraph{\textbf{Game Mechanisms and Narratives}}
Most scenarios applied some form of narratives and game mechanics to bring the user into the context of VR. Using enriched narratives is associated with increased presence \cite{weechNarrativeGamingExperience2020}. Current gaming applications in VR have already ``remote-controlled'' the VR user's physical actions through the game design. For example, VR rhythm games make the user do dancing poses originating from the song \cite{vrfitnessinsiderBeatSaberGANGNAM2018}, or players have to maintain different poses by putting their head and hands in the right spot, which can be dabbing, lunges, squats, or even choreography (e.g., OhShape \cite{labOhShapeNewVR2020}). Because the VR user is immersed in the game and unaware of what their physical actions represent in the real world, malicious actor can make them do inappropriate posture to confront bystanders as described in \textit{Insult Simulator}.

\paragraph{\textbf{Habituation and Trust}}
In a discussion during W1, P2 mentioned, \textit{``because I believe any application we are talking about right now here requires a degree of trust.''} This trust in a VR application (system) can be built by the habituation to the environment or interaction. An example would be \textit{Falsely Mapped Apartment} where the VR user is used to a fully mapped place. Malicious attacks remove or add a virtual object at one point to break this habituation and trust in the system. Another example is \textit{Moving Platform} where the user interacts with a haptic display, and suddenly the system stops (accidentally or deliberately) to provoke physical harm. The VR user gets used to the interaction and is fully committed to the action they are doing. Then comes the moment to break the habituation and provoke physical harm.

\section{Demonstrating the Potential for VPPM harm}
The workshop illustrates the significant scope and scale of physical harm potentially enabled by {\VPPM}s. However, it would be easy to write off many of these attacks as infeasible or impractical since our main method was grounded in speculative design and workshops. 

To demonstrate that the potential for physical harm related to {\VPPM}s is both plausible and pressing, we introduce two implementations of \VPPM\ concepts. These implementations are grounded in two recent publications from CHI \cite{azmandianHapticRetargetingDynamic2016,rietzlerBreakingTrackingEnabling2018}. We deliberately choose two publications from our community to emphasize the responsibility we carry when creating such techniques. Our two implementations are meant to demonstrate that with the information from the paper and some basic computer science knowledge, we were able to create two applications that are using the \emph{puppetry} and \emph{mismatching} attacks. These two applications could potentially be uploaded to open stores such as SideQuest and cause a certain amount of harm to the current early adopter population of VR technology. While they could be counteracted with simple additions to the publication process or platform-level mitigation (see section \ref{sec:mitigations}, Mitigations and Countermeasures), these are currently not in place. The existence of these current weak spots should be an additional call to action to platform developers and markets. Two applications (\SteppingOn\ and \HittingFace) are mainly leveraging the predominant form of \VPPM\ (\emph{puppetry} attack) exemplifying how {\VPPM}s can be easily subverted and provoke physical harm to the VR user.

\begin{figure}[t]
\begin{center}
  \begin{tabular}{@{\hspace{0.1cm}}c}
		\includegraphics[width=\linewidth]{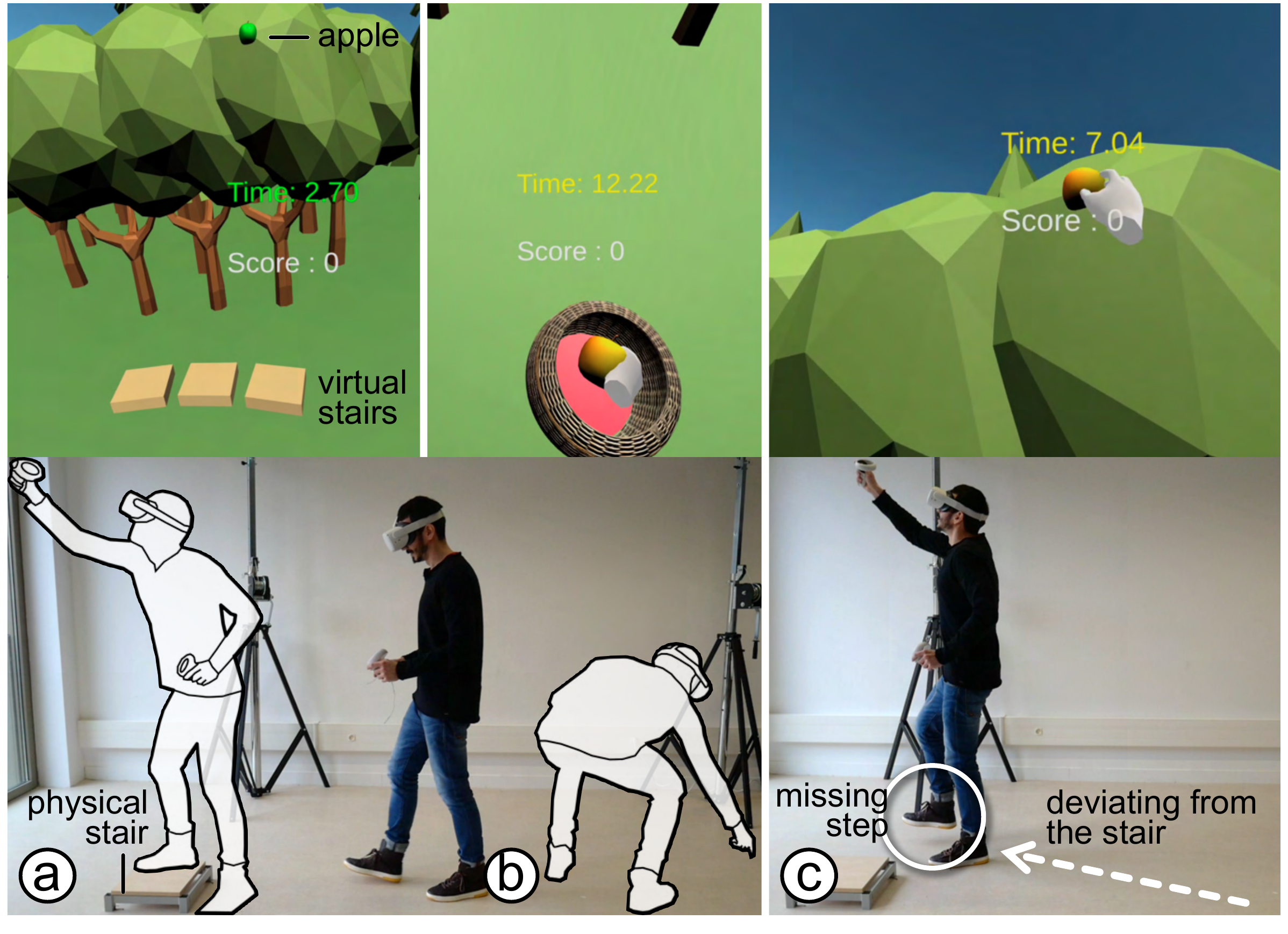}
  \end{tabular}
  \caption{(a) The SteppingOn setup has one physical and three virtual stairs. The application redirects the VR user to match the stepping feedback on the physical stair while (b) walking back and forth for collecting apples. (c) The application randomly turns off redirection to create a missing step.}
  \Description[Figure 6 illustrates the application, SteppingOn, that redirects a VR user towards a missing step, provoking falling over.]{Figure 6a shows the setup of SteppingOn. A VR user is redirected towards the same physical stair while stepping on three different virtual stairs to collect apples as fast and as many as possible. Figure 6b shows the application applies redirection technique when the VR user turns back to put the apple inside a basket. Figure 6c illustrates that at one point the application stops the redirected walking so that the user deviates from the physical stair and provokes a missing step.}
  \label{fig:app-steppingon}
\end{center}
\end{figure}




\subsection{SteppingOn: Provoking Missing Steps Using Redirected Walking}

\SteppingOn\ enables the haptic feedback of stepping on a stair to collect virtual items in VR. The setup (Figure \ref{fig:app-steppingon}a) contains one physical stair functioning as a prop in the real world to support the haptic feedback of three virtual stairs in VR. The user has to walk towards the three virtual stairs to pick apples from the trees and return to the original point to put the apple at a certain position in VR (Figure \ref{fig:app-steppingon}b). \SteppingOn\ always redirects the user toward the same physical stair while having the impression of visiting a different virtual stair each time. When the user drops an apple and turns their head to go back to the stairs, we rotate the VR scene. The rotation of the scene is imperceptible. Once the virtual stair aligns with the physical one, we stop rotating to prevent the alignment from being exceeded. Finally, we add two game mechanics (score and time limit) to make the user commit to grab the apples and climb the stairs. The user must collect as many apples and as fast as they can. 

During the game, the application randomly turns off the redirection so that the user deviates from the targeted physical stair and makes a missing step (Figure \ref{fig:app-steppingon}c). This effect is similar to the moment when climbing stairs, where we think there is one more tread, but we are already standing at the landing, therefore, making an additional step. The missing step effect sometimes triggers small forms of a stumble and can be easily increased using a higher stair.  The setup was inspired by Haptic Retargeting \cite{azmandianHapticRetargetingDynamic2016} and the concept of redirected walking \cite{razzaqueRedirectedWalking2001}. 

\begin{figure}[t]
\begin{center}
  \begin{tabular}{@{\hspace{0.1cm}}c}
		\includegraphics[width=\linewidth]{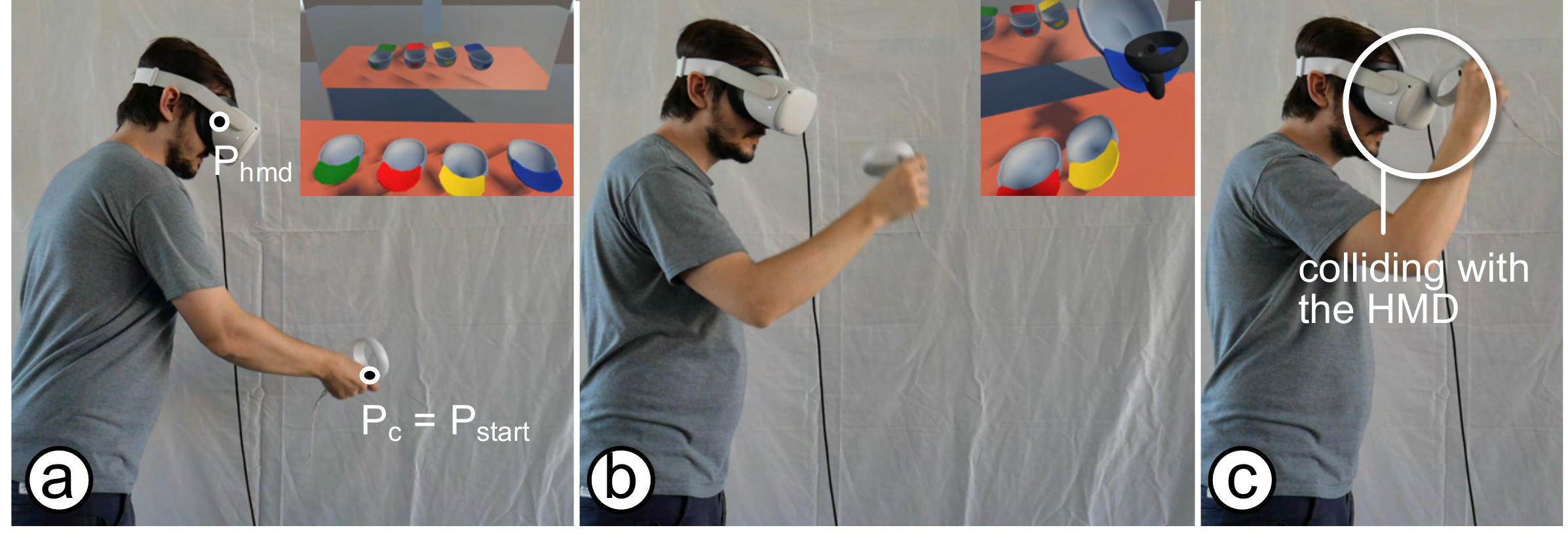}
  \end{tabular}
  \caption{(a) A VR user tests several baseball caps on his avatar in VR. (b) The concept of HittingFace is to change the offset between the virtual and physical movements while the user moving the controller closer to the HMD. (c) Because of the trajectory of the controller changes during the movement, HittingFace is able to provoke collision.}
  \Description[Figure 7 shows the concept of HittingFace, where the application increases the offset between the virtual and physical position of the controller, provoking a collision with the headset.]{Figure 7a shows a VR user is testing several baseball caps. Figure 7b shows that the application changes the offset between the virtual and physical movements while the user moving the controller closer to the headset. Figure 7c demonstrates that the trajectory of the controller changes abruptly during the movement. The VR user follows the visual in VR, moving their controller closer towards the headset and provoking a collision.}
  \label{fig:app-hittingface}
\end{center}
\end{figure}

\subsection{HittingFace: Changing the Trajectory of Controller Movement  to Provoke Collision with the HMD}
\HittingFace\ is a short example application that manipulates the trajectory of hands by adding an offset between the virtual and physical position of the controller to provoke collision between the controller and HMD. Figure \ref{fig:app-hittingface}a shows the scenario of \HittingFace\ where the VR user puts on different baseball caps on their avatar to test their outfit. When the user selects a cap with controller, the application records the controller position ($P_{c}$) as the starting position ($P_{start}$). Next, we calculate $(P_{hmd} - P_c) / (P_{hmd} - P_{start})$ as an indicator of how close the controller and the headset are. When the VR user puts on the baseball cap in VR, the controller is closer to the HMD. We add an offset to the the direction of facing-forward. The application increases the offset abruptly, shifting the visual of controller away from the real one. Then the VR user moves the controller even closer to the HMD (Figure \ref{fig:app-hittingface}b), provoking collision in Figure \ref{fig:app-hittingface}c. This application was inspired by Breaking the Tracking \cite{rietzlerBreakingTrackingEnabling2018} that simulates the feedback of weight in VR by using perceptible tracking offsets.

\subsection{Reflection}
We started by defining the physical harm we wanted to provoke (e.g., fall, collision with the HMD). Next, we were thinking about incorporating physical harm into the physical movements and some game mechanics. Inspired by habituation, the applications exploit the manipulations after the user becomes familiar with the interaction. At one point, the applications start to nudge the user's physical movement (e.g., walking direction, hand movement trajectory) and provoke the physical harm we chose. 
Implementing these sample applications (\SteppingOn\ and \HittingFace) shows how current concepts from \VPPM\ research can be trivially subverted. 

We did not evaluate both applications due to the high risk of hurting participants. The implication of presenting both applications is to show how easy it can be to subvert an existing \VPPM\ to provoke physical harm. Both demonstrations may seem easy to counter. An example would be detecting the discrepancy between the virtual and physical movements as a threshold to stop a \VPPM\ technique. However, the malicious use of {\VPPM}s and its countermeasure are both unexplored spaces for researchers and practitioners currently. The goal of the two applications is to raise awareness and initiate discussions in the HCI and VR communities. We further discuss mitigations and preventative recommendations for the malicious use of {\VPPM}s from the end-user to the platform level in section \ref{sec:mitigations}.



\section{Mitigations and Countermeasures}
\label{sec:mitigations}
We have discussed the potential attacks, physical harm, and how to provoke them using {\VPPM}s. In this section, we reflect on mitigations and preventative recommendations against the malicious use of {\VPPM}s for practitioners and researchers.

\paragraph{\textbf{Awareness and Consent of VR Users}}
When applying {\VPPM}s, the user may be subjected to manipulation knowingly or unknowingly, and the manipulation may or may not be perceptible to the user. This notion is one possibility of how malicious actors hide their intention and provoke physical harm to the VR user. Given this, it would be reasonable to suggest future VR applications using {\VPPM}s should at-a-minimum disclose that such an approach is being used and particularly the \textit{intent} behind its usage. 

Where a \VPPM\ might be particularly risky or open to abuse, we would suggest it should be described to the user in sufficient detail to seek informed consent for applying such manipulations and perceptual hacks. For example, applications should be transparent about what kinds of actions are manipulated using \VPPM, how these actions are represented, and the possible effects on VR users \cite{breyEthicsRepresentationAction1999a, breyVirtualRealityComputer2014}. At the same time, VR users are freely able to select different levels of deception provided by {\VPPM}s \cite{slaterEthicsRealismVirtual2020}. This concept originates from reducing the realism of an XR application if a user only wants to try a little taste of the virtual environment. By providing this option, VR users can voluntarily choose to what degree they want to be manipulated by {\VPPM}s if they feel comfortable with the manipulation. Applications using {\VPPM}s also need to respect the VR user's right to withdraw anytime by providing an opt-out option for stopping the \VPPM\ technique \cite{benfordUncomfortableInteractions2012, slaterEthicsRealismVirtual2020}.

\paragraph{\textbf{Validation / App Store Protections}}
App platforms (e.g., Steam, Oculus Store, SideQuest) also need to verify what type of and how much \VPPM\ is used in an application. In the same way that malicious actors have access to reference implementations and perceptual thresholds, so do the platforms that profit off of selling XR applications and experiences. Thus we assert the responsibility should, in part, fall on their shoulders to seek out ways to detect the presence of such manipulations in applications that they provide. In the long run, these platforms should build a standardized rating system for induced contents \cite{wilsonViolentVideoGames2018a, spiegelEthicsVirtualReality2018} and {\VPPM}s as additional information for end-users.

\paragraph{\textbf{Platform-Level Mitigations: Provision and Detection}}
We anticipate that platform-level APIs (e.g., OpenXR\footnote{\url{https://www.khronos.org/openxr/}}) could provide access to safe, permitted, and validated {\VPPM}s that tie into mechanisms for awareness and consent. An example would be an OpenXR software library of redirection techniques that could prevent malicious implementations. 

Considering the pipeline of AR/VR technology, a device requires sensing the raw data, extract information for the recognition of high-level semantics, and rendering on top of the HMD \cite{roesnerSecurityPrivacyAugmented2014}. Platforms could implement low-level protections against the unpermitted usage of {\VPPM}s in the sensing and rendering. For example, the discrepancy between virtual and physical movements could be monitored \cite{lebeck2016safely, lebeckSecuringAugmentedReality2017a}. If the physical movement deviates significantly from the virtual movement, this could reveal some types of {\VPPM} (e.g., the gain-type ones). Similarly, one could imagine the platform detects the dangerous overlap between virtual contents and the physical environment. An example would be a virtual target overlaid on a physical lamp, which sounds like a risk of non-organism damage. This type of mitigation can be a part of reality-aware headsets where the virtual and physical context needs to be considered in making the experience safer for users.

Lastly, on the device level, one can apply permission-based security with access control lists \cite{barreraMethodologyEmpiricalAnalysis2010}. Therefore a VR system may prevent a malicious third-party application from abusing access to the sensory data. For example, blocking the access to the captured image of cameras to avoid incorporating bystanders as an enemy avatar in \textit{Start A Fight}.

\paragraph{\textbf{Community-led Regulations and Guidelines}}
In time, we would expect that regulations could be formed around our proposed mitigations and preventative measures. There are a number of routes that could accomplish this. Most immediately, we propose such regulation could be formulated by not-for-profit organizations in this space (e.g., XRSI\footnote{\url{https://xrsi.org/}}), creating voluntary guidelines that could guide the actions both of app platforms and app developers \cite{PearlmanTalkAtUSENIX}. Eventually, one could imagine firmer legal protections being put in place. An example would be an equivalent of GDPR\footnote{\url{https://gdpr-info.eu/}} such as an extended reality protection regulation (XRPR) that would include the \textit{right to perceptual integrity}. As recent works discussed on human rights of neurotechnology (e.g., \cite{yuste2017four, yuste2021s}, and The NeuroRights Foundation\footnote{\url{https://neurorightsfoundation.org/}}), XRPR also has to include the right to agency and consent to choose one's own actions while using {\VPPM}s.

\paragraph{\textbf{The Role of the Research Community: Anticipation and Disclosure}}
{\VPPM}s offer obvious advantages to interaction design and locomotion in particular, having been repeatedly pursued by research. Consequently, the implementation details of {\VPPM}s and the perception thresholds found are open to everyone. However, this information is also available to malicious actors. This early insight gives malicious actors the chance to abusively exploit published results and concepts, for example, using {\VPPM}s to enact harmful consequences on VR users. Fundamentally, the current way we apply and publish {\VPPM}s is hacking human perception. We consider this hack as exposing a weak spot of our perceptual vulnerabilities. One can provide patches to fix the software backdoor, as we have previously discussed, but there is no patch to fix the hack of our perception directly.

In our view, these risks necessitate a change of approach regarding how we disseminate novel research related to {\VPPM}s. We suggest that the research community should publish \VPPM\ with the potential threats/risks in mind. The community should consider the perceptibility of a given \VPPM\ instead of only optimizing for presence, immersion, and other usability measurements (e.g., performance). This approach would ensure one could apply \VPPM\ always above the perception threshold during VR interaction, allowing VR users to know they are interacting with a certain degree of manipulation. This idea is already starting to get explored in the field of locomotion. Rietzler \textit{et al.} \cite{rietzlerRethinkingRedirectedWalking2018} proposed using perceptible thresholds to reduce the space requirement for redirected walking, which could also benefit a transparent usage of \VPPM.
Finally, we suggest if a \VPPM\ publication has the potential to enable abusive outcomes (e.g., if it has the potential to facilitate one of the attacks identified herein), then the author(s) should include discussion regarding the potential threat/risk posed at-a-minimum.

\section{Discussion}
Our goal with this paper was to start the first exploration into how dangerous current {\VPPM}s could become in the future. While we are able to observe current applications of {\VPPM}s, we needed to apply speculative design methods to try to predict how these current {\VPPM}s could be subverted in the future. Applying this method allowed us to present a definition of {\VPPM}s and a set of speculative scenarios which we used to derive a classification of attacks and gain a better understanding of the characteristics of the potential harm arising from the {\VPPM}s. We identified five potential attacks (\emph{puppetry}: walking, arm-movement; \emph{mismatching}: false-positive, false-negative, swapping), a categorization of harm (provoke/receive matrix), and two variables that participants found particularly important when thinking about {\VPPM}s (\textit{severity of physical harm} and \textit{perceived agency}).

Physical harm is a novel problem that arises at the intersection of HCI, XR, and Security/Safety research. This unique combination aims at using methods from security research, combined with insights from HCI which are then applied to applications in XR. Additionally, XR may become an ``ideal'' platform to abuse perceptual vulnerabilities and manipulate the user's motion. The ability to manipulate the VR user's physical movements and actions could have way more impact than only hitting a piece of furniture.

Pursuing positive outcomes (e.g., speed, accuracy, enjoyment) is usually a common goal for HCI research. {\VPPM}s help in overcoming the limitations of VR technologies. They also expose a weak spot of VR users who are particularly vulnerable because of losing connection with the real world. 
Our intention is to raise the awareness that the interaction design in VR using {\VPPM}s could be used for malicious intention as well. Although examples shown in the applications may be easily thwarted, this is currently not the case because {\VPPM}s are mostly used inside research. Meanwhile, it is necessary to ensure that developers are aware of these potential attacks and that they take measures to prevent or mitigate them. We want to emphasize the importance of the safety and security of the VR user, with a particular focus on physical harm done by human perception hacking. To our best knowledge, we are the first to establish the term, organize the knowledge in this domain, and lay out suggestions on how to deal with {\VPPM}s (i.e., section \ref{sec:mitigations}).

\subsection{Limitations}
Our work encounters a methodological situation known as the Collingridge dilemma\footnote{\url{https://en.wikipedia.org/wiki/Collingridge_dilemma}}. The malicious use of {\VPPM}s cannot be easily predicted until they are extensively developed and widely used. However, at the point we can do that, the control or change to affect the usage of {\VPPM}s is difficult because the technology has become entrenched. Therefore, we chose speculative design as our approach to both critique current practices, and reflect on future technologies and their implications.

The resulting scenarios show the possibilities of potential harm exploited by {\VPPM}s. Using a speculative design workshop allows us to broadly explore this space. However, one outcome that we are not able to assess with the current method is the likelihood of malicious attacks using {\VPPM}s and the occurrence of physical harm in the everyday usage of VR.
Nonetheless, surveying the in-the-wild VR phenomena (e.g., VR fails \cite{daoBadBreakdownsUseful2021} or interactions between VR users and bystanders \cite{ismar2021ohagan}) could provide one route towards early detection of these attacks happening in practice, and such research would be aided by our findings.


Our participants were from HCI research and design research background. The resulting scenarios were more interaction design research oriented. We did not interpret the results depending on the participant’s expertise because participants collaborated during the workshop to create outcomes (scenarios, dimensions). {\VPPM}s are mainly used inside research currently. Therefore inputs from our participants are valid because it reflects on how research communities perceive the malicious use of \VPPM\ and how we can mitigate it in the future. However, we acknowledge that our current results show only one perspective of the malicious use of {\VPPM}s. Future research should consider similar studies and experiments with people from the safety and security area, technical VR/XR, and dark design patterns to provide in-depth technical details in this direction.

\subsection{Future Work}
Our work is a first exploration into a topic that could potentially grow exponentially in its risk at the moment when we have always-on XR devices. Based on our current findings, we open the door to further research into the malicious potential of perceptually manipulating users in the context of XR.

\paragraph{\textbf{Intent beyond Physical Harm}}
Currently, this paper focus on the physical harm, but we want to point out that the malicious user of {\VPPM}s could accomplish more than `just' provoking physical harm. The realism of VR technology can induce certain behavioral changes (e.g., given the virtual representation in VR, users with taller avatars negotiated more aggressively than users with shorter avatars \cite{yeeProteusEffectImplications2009}). Slater and colleagues \cite{slaterEthicsRealismVirtual2020} discussed the psychological realism of AR/VR and its possible impact on the user. In both workshops, participants (P2 and P5) mentioned the possibility of exploiting psychological harm to the user (e.g., VR application introduces a phobia to the VR user and make them forever be afraid of using an HMD). Unlike perceptually manipulating the physical movements, the psychological harm cannot only provoke immediate effect and reaction but also the long-term impact (e.g., trauma or phobia). 

\paragraph{\textbf{Harm beyond the VR User, and the Here-and-Now}}
Although we focus on provoking harm to one VR user, malicious attacks could easily go beyond that. We already find some examples in our workshops. For instance, \textit{Start a Fight} (S07) renders bystanders as enemies in VR and makes the VR user punch them or vice versa. The other example is \textit{Minecraftish} (S12) that the VR user throws an object at pedestrians. In the results of synthesizing step, P3 presented the social involvement dimension (D5) that starts with ``harm yourself'' to ``harm others''. Harm others could be exploited in several ways such as hitting bystanders (S07), insulting people (S11), or let others watch the VR user suffering or even dying (S04 and S09). The target of malicious actors varies from a VR user, multiple VR users, bystanders, to objects and organisms in the environment.  {\VPPM}s could also be used to create the circumstances for harm in the future, e.g., using the VR user to manipulate elements in the physical environment that might cause harm to bystanders later. We have examined only a narrow scope of the potential harms that could be made possible by  {\VPPM}s in the future, and suggest consideration be given to further understanding multi-user {\VPPM}s, harm beyond the VR user, and creating the circumstances for harm beyond the VR session.

\paragraph{\textbf{Challenges of VPPMs in AR and XR}}
We anticipate that researchers and practitioners can also apply {\VPPM}s to AR and XR in the future. As an example, Optical Marionette \cite{ishiiOpticalMarionetteGraphical2016} applied redirected walking on video see-through HMDs. In video see-through HMDs, malicious actors are still able to apply both puppetry and mismatching attacks since they still have full control over the visuals of the user. However, when using optical see-through HMDs (e.g., deceptive holograms \cite{8418615}), applying puppetry attacks becomes more challenging because the user can observe their physical movements at the same time. 
Future VR, AR, and XR technologies would allow the user to break free the static play space towards moving around freely in the world. The safety risk may be amplified, and \textit{mismatching} attacks are still able to trick the user (e.g., substitute the virtual and physical content on video/optical see-through HMDs to provoke falling over). 
Future research could continue to explore the novel attacks using {\VPPM}s in this direction, understanding the common attacks shared across XR devices.

Broadly, whilst it would be understandable if there was still some scepticism regarding the prescience of the risks posed by {\VPPM}s, it is our view that we have only just begun to understand the extent to which XR users are exposed to risks through these techniques. As XR technology and its requisite sensing grow in capability, so too will a malicious actors ability to exploit this technology for harmful intent. Consequently, it is paramount that research to this end be considered and acted upon before real harm is inflicted upon real users.
\section{Conclusion}
In this paper, we define \VPPM\ as XR-driven exploits that alter the human multi-sensory perception of our physical actions and reactions to nudge the user's physical movements. 
Through speculative design workshops, we collect a set of harmful scenarios using {\VPPM}s, identify two main classes (\emph{puppetry} and \emph{mismatching}) of potential attacks, and characterize physical harm.
Two sample applications (\SteppingOn\ and \HittingFace) are implemented as an demonstration to show how current concepts from \VPPM\ research can be trivially subverted.
Finally, we propose platform-level mitigations and preventative recommendations for practitioners and researchers against the malicious use of {\VPPM}s. Our work opens new research directions at the intersection between HCI, XR, and security research.
We want to raise awareness that the current way we apply and publish {\VPPM}s can lead to malicious use of our perceptual vulnerabilities. We consider the current practice provides a dangerous leak of human perceptual weak spots --- human perception thresholds that cannot be patched --- which can be used by future malicious actors. 
Overall, we argue that {\VPPM}s do have the potential to be misused to provoke physical harm in the future and HCI as an academic discipline should become more cautious publishing such work and also reflect on the potential for abuse. 

\begin{acks}
This work was partially conducted within the HARMFULVR JCJC project (ANR-21-CE33-0013) funded by French National Research Agency (ANR). We appreciate all the anonymous reviewers for their advice to improve this paper, and we thank the workshop participants for their contributions and time.
\end{acks}

\appendix
\section{Descriptions of Selected Scenarios}
\label{apx:scenarios}
\paragraph{[Magic Maze (S03), exploits: redirected walking, physical harm: fall]}
\textit{Magic Maze} is an application where the VR user explores a virtual maze in their apartment or a building. The application applies redirected walking to the VR user to control their walking direction in this space. As the application steers the VR user towards a stairway, they are unaware of the height difference and fall.

 \paragraph{[Start a Fight (S07), exploits: swapping, physical harm: punch other, get punched]}
In this scenario, a VR user plays a game in a public space where the goal is to fight enemies. The VR application detects bystanders in the real world and maps the enemy's avatar onto bystander so that the VR user punches them. This would result in harm to bystanders and potential harm for the users. 

\paragraph{[Getting Robbed (S08), exploits: redirected walking, physical harm: get robbed, stabbed]}
\textit{Getting Robbed} shows that a VR user is in a game like Pokemon Go and needs to walk around in VR to collect items. The items are located in dangerous places in the real world, and all the physical surroundings are replaced by the virtual game view. This results in a practically blindfolded user walking and not knowing where they are headed. This will then be abused once the victim was lured into a dangerous area (e.g., being robbed or physically attacked in an alley). 

\paragraph{[Catch a Ride (S09), exploits: redirected walking, overplay audio feedback, physical harm: get driven over]}
\textit{Catch A Ride} is where a VR user is immersed in a VR game at an open space. The game has loud audio feedback that can overplay the sound from the real world. The game redirects the user onto an open road so they get hit by a car since they are not able to see or hear the traffic noise.

\paragraph{[Falsely Mapped Apartment (S10), exploits: remove or add virtual object in an one-to-one mapped environment, physical harm: fall or collision]}
In this scenario, a future VR technology allows users to re-create a fully-mapped apartment in VR. The VR user can touch anything and sit  anywhere as they do in the real world, believing that this mapping matches their real world home. The user habituates to this environment as each real-world object is mapped to a VR one. Malicious actors may exploit this by adding or removing virtual objects. Adding VR objects may result in injuring the user by, for example, sitting on a VR chair that has no physical counterpart. Similarly, removing VR objects may result in collisions with real-world objects, such as real-world tables that do not have counterparts in VR. The idea in this scenario is to first make accustomed to having a one-to-one mapping between the virtual and the real world and trust that this is the case, and then  introduce/remove objects to unexpectedly break this mapping. 

\paragraph{[Insult Simulator (S11), exploits: game mechanics, physical harm: insult bystanders, get punched]}
In this scenario, the VR user plays a game in an open space where people are around, and the game mechanics lead the user to perform physical actions that appear insulting for onlookers without enough context. In an illustrated example by our participants, a user follows the narratives in VR to reach out with bare hands but may seem like they are performing a Nazi salute from outside. A bystander that does not know what the VR user is doing may feel insulted/offended as a result. 

\paragraph{[Minecraftish (S12), exploits: redirected haptics and swapping, physical harm: throwing objects at others]}
\textit{Minecraftish} is another scenario that uses redirected haptics to make the VR user grab a real-world object that they think resembles a counter part in a Minecraft-style VR game. The application can access information captured by the VR headset, and at some point, it redirects the VR user to grab an object (or a pet) resembling the virtual content in the environment. Because the VR user thinks they are doing the task in VR and do not perceive the difference, they stack up or even throw a potentially harmful object (e.g., hot drink or sharp object) outside the window and hit pedestrians. The physical harm in this scenario affects the personal objects or other organisms (e.g., pet, bystander) in the environment.

\bibliographystyle{ACM-Reference-Format}
\bibliography{main}










\end{document}